\documentclass{PoS}

\title{Experimental determination of $V_{us}$ from kaon decays}
\ShortTitle{$V_{us}$ from kaon decays}

\author{\speaker{Matthew Moulson}\\
        INFN Laboratori Nazionali di Frascati, Italy\\
        E-mail: \email{moulson@lnf.infn.it}}

\abstract{
  The status of the experimental determination of $V_{us}$ from
  $K_{\ell3}$ and $K_{\mu2}$ decays is reviewed.
  Factors currently limiting the precision of the evaluation
  of $V_{us}$ from kaon decays and prospects for new measurements
  within the next few years are discussed.}

\FullConference{9th International Workshop on the CKM Unitarity Triangle\\
		28  November - 3 December 2016\\
		Tata Institute for Fundamental Research (TIFR), Mumbai, India}

\begin{document}

\section{Introduction}

At present, the most stringent test of CKM unitarity is obtained from the
first-row condition $V_{ud}^2+V_{us}^2+V_{ub}^2 = 1 + \Delta_{\rm CKM}$.
Between 2003 and 2010, a wealth of new measurements of
$K_{\ell3}$ and $K_{\ell2}$ decays, together with steady theoretical
progress, made possible precision tests of the Standard Model
based on this relation. In a 2010 review, the FlaviaNet
Working Group on Kaon Decays set bounds on $\Delta_{\rm CKM}$ at the level
of 0.1\%~\cite{Antonelli:2010yf}. The 2010 FlaviaNet
evaluation was most recently updated in 2014~\cite{Moulson:2014cra}.
Since the 2014 update, experimental progress has been minor, but some
issues have come to the fore that illustrate both the potential for future
progress and the obstacles to overcome.

The determination of $V_{us}$ from $K_{\ell3}$ and $K_{\ell2}$ decays requires
values for the hadronic constants $f_+(0)$ and $f_K/f_\pi$, respectively.
At the moment, the uncertainties on the lattice
QCD values for these constants contribute slightly more than those for the
experimental data to the overall uncertainty on $V_{us}$.
In recent years, however, the pace of theoretical progress has exceeded
that of experiment. Advances in algorithmic sophistication and
computing power have lead to a number of new lattice QCD results with
total uncertainties at the level of 0.3\%.
The recently released edition of the biannual review from the
Flavor Lattice Averaging Group (FLAG) provides a critical overview and
recommended values of the lattice constants entering into the evaluation
of $V_{us}$~\cite{Aoki:2016frl}. Progress on lattice results for the
evaluation of $V_{us}$ was also reviewed at this
conference~\cite{Simula:2017brd}.

\section{Experimental inputs}

\subsection{Branching ratios and lifetimes}

Existing branching ratio (BR) and lifetime measurements allow $V_{us}$
to be evaluated from the rates for the $K_{e3}$ decays of the
$K_S$, $K_L$, and $K^\pm$ and for the $K_{\mu3}$ decays of the $K_L$ and $K^+$;
the rate for $K^\pm_{\mu2}$ provides an independent evaluation of
$V_{us}/V_{ud}$.
Most of the available data is in the form of ratios of BRs, and
the few absolute BR measurements in the data set have
residual dependence on the kaon lifetimes via the experimental acceptance.
Therefore, the best values for the leptonic and semileptonic decay rates
are obtained from fits to the measured values of the lifetimes and of
all of the BRs for the major decay modes, with the sum of the BRs
constrained.
As a result, new BR measurements of any of the major decay modes
are potentially interesting inputs to the analysis, not just the
$K_{\ell3}$ and $K^\pm_{\mu2}$ modes.

There have been a handful of new measurements entering into the
fits since 2010, but none since the 2014 update. The most significant
recent development is the 2014 measurement of
${\rm BR}(K^\pm\to\pi^\pm\pi^+\pi^-)$ from KLOE-2~\cite{Babusci:2014hxa}
which filled a significant gap in the $K^\pm$ data set---there was very
little previous constraint on this BR. 
The inclusion of this measurement significantly reduced the uncertainty
on the fit result for ${\rm BR}(K^\pm_{\mu2})$.

\begin{table}
\begin{center}
\begin{tabular}{lccccc}
\hline\hline
Parameter & Value & $S$ & \multicolumn{3}{c}{Correlation coeff. (\%)} \\
\hline
$K_S$ \\
\hspace{1em}${\rm BR}(K_{e3})$ & $7.05(8)\times10^{-4}$ \\
\hspace{1em}${\tau_{K_S}}$ & 89.58(4)~{\rm ps} \\ \hline
$K_L$ \\
\hspace{1em}${\rm BR}(K_{e3})$ & $40.56(9)\%$ & 1.3 \\
\hspace{1em}${\rm BR}(K_{\mu3})$ & $27.04(10)\%$ & 1.5 & $-28$ \\
\hspace{1em}${\tau_{K_L}}$ & 51.16(21)~{\rm ns} & 1.1 & $+8$ & $+14$ \\ \hline
$K^\pm$ \\
\hspace{1em}${\rm BR}(K_{\mu2})$ & $63.58(11)\%$ & 1.1 \\
\hspace{1em}${\rm BR}(K_{e3})$ & $5.088(27)\%$ & 1.2 & $-68$ \\
\hspace{1em}${\rm BR}(K_{\mu3})$ & $3.366(30)\%$ & 1.9 & $-52$ & $+38$ \\
\hspace{1em}${\tau_{K^\pm}}$ & 12.384(15)~{\rm ns} & 1.2 & $+5$ & $+1$ & $0$ \\
\hline\hline
\end{tabular}
\end{center}
\caption{\label{tab:br}
  Summary of branching ratio and lifetime measurements from the fits
  to $K_S$, $K_L$, and $K^\pm$ BR and lifetime measures.
  Scale factors are calculated using the PDG prescription~\cite{Olive:2016xmw}.}
\end{table}

Table~\ref{tab:br} summarizes the results of the BR and lifetime fits used
for the analysis. The input data sets for the $K_L$ and especially
the $K^+$ contain some inconsistent measurements, and the fits have
$\chi^2/{\rm ndf} = 19.8/12$ ($P=7.0\%$) and
$\chi^2/{\rm ndf} = 25.5/11$ ($P=0.78\%$), respectively. 
These fits are more selective in the use of older measurements than the fits
by the PDG~\cite{Olive:2016xmw}, and there are some differences in the
handling of correlations and dependence on external parameters.
Relative to the PDG fits, the results of the $K^\pm$ fit for the
$K_{\ell3}$ BRs have central values that are 0.3--0.4\% higher and
slightly smaller uncertainties.
The PDG fits for the $K_L$ and $K^+$ have
$\chi^2/{\rm ndf} = 37.4/17$ ($P=0.30\%$) and
$\chi^2/{\rm ndf} = 53/28$ ($P=0.26\%$), respectively. 

\subsection{Form-factor parameters}
\label{sec:ff}

\begin{figure}
\begin{center}
\includegraphics[height=0.35\textheight]{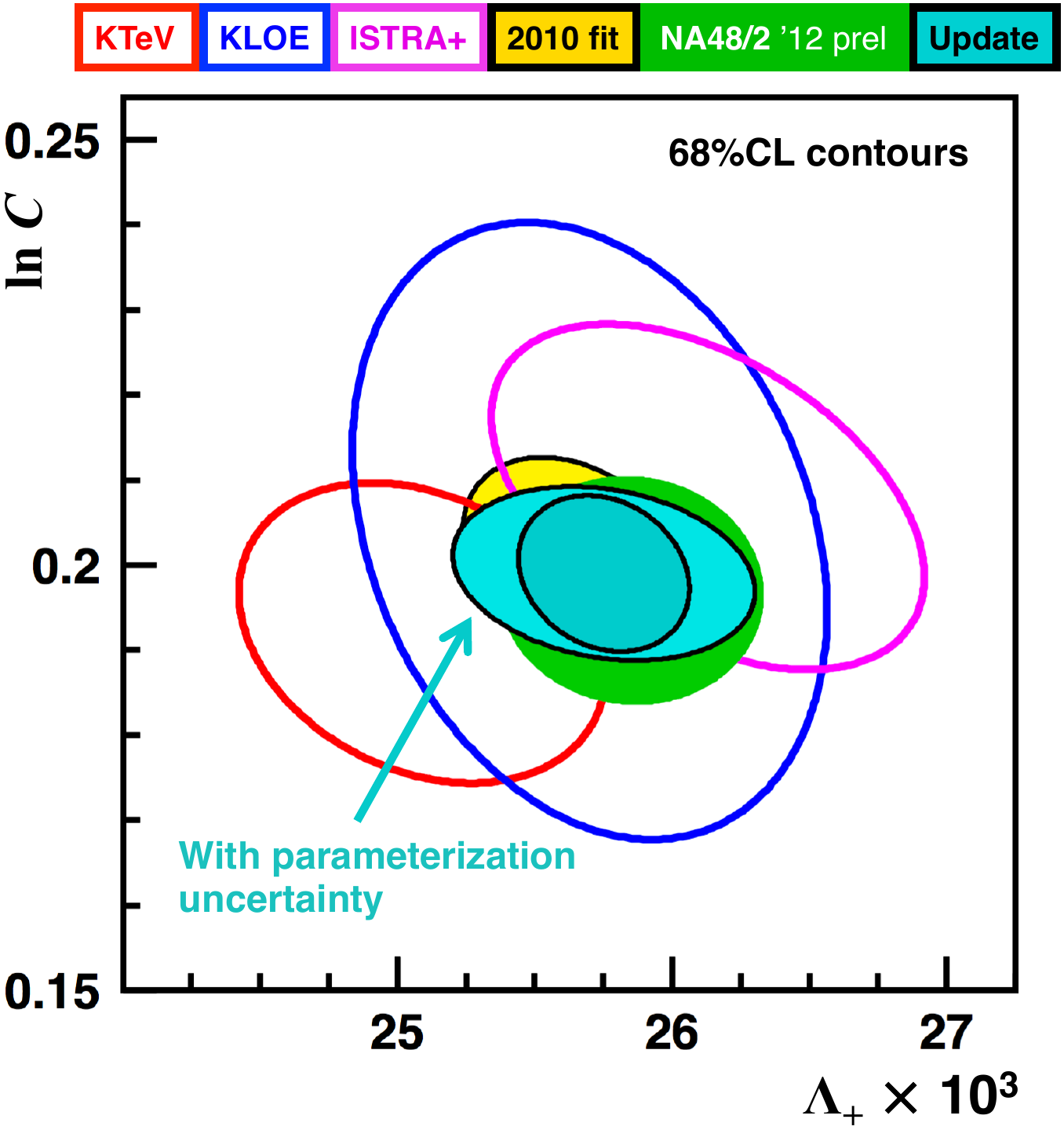}
\caption{68\% confidence contours for form-factor parameters 
  from dispersive fits, for different experiments
  ($K_{e3}$-$K_{\mu3}$ averages)
  The FlaviaNet 2010 average and the updated average,
  with the NA48/2 result included, are shown. The confidence contours
  do not include the contribution to the uncertainties from
  the dispersive parameterization, except in the case of the outer
  ellipse for the updated fit.}
\label{fig:ffcomp}
\end{center}
\end{figure} 
The 2010 analysis included measurements of the $K_{\ell3}$
form-factor parameters for $K_L$ decays from KLOE, KTeV, and NA48, and
for $K^-$ decays from ISTRA+. Because of the advantages
described in~\cite{Antonelli:2010yf}, the results from fits using the
dispersive representation of~\cite{Bernard:2009zm}
were used to evaluate the phase-space integrals.
In the dispersive representation, the vector and scalar form factors
are each specified by a single parameter: $\Lambda_+$, which is essentially
the slope of the vector form factor at $t=0$, and $\ln C$, which is
related to the value of the scalar form factor at the unphysical
Callan-Treiman point, $t=m_K^2-m_\pi^2$.
In 2012, the NA48/2 experiment released preliminary results for the 
form factors for $K^\pm_{\ell3}$ decays
obtained with the quadratic-linear
$(\lambda'_+, \lambda''_+, \lambda_0)$ and polar $(M_V, M_S)$
parameterizations~\cite{Wanke:2012xk}.
However, it is possible to obtain approximately equivalent values
of $\Lambda_+$ and $\ln C$ from a fit to the NA48/2 measurements of
$(\lambda'_+, \lambda''_+, \lambda_0)$ using the expressions in the appendix 
of~\cite{Bernard:2009zm} and the observation that 
$\lambda_0 \approx \lambda_0' + 3.5\lambda_0''$~\cite{Ambrosino:2007ad}.
Including the preliminary NA48/2 results, appropriately converted for the
current purposes, the dispersive average becomes
$\Lambda_+ = (25.75\pm0.36)\times10^{-3}$, 
$\ln C = 0.1985(70)$, with $\rho = -0.202$ and $P(\chi^2) = 55\%$
(see Fig~\ref{fig:ffcomp}).
The central values of the phase-space integrals barely change with this 
inclusion; the uncertainties are reduced by 20\%.

The above situation is unchanged from 2014.
In 2016, the OKA experiment, the successor to ISTRA+,
updated its preliminary results on the $K_{e3}$
form-factor parameters from polynomial, pole, and dispersive fits, obtained
with a sample of 3.2M $K^+$ decays \cite{Obraztsov:2017oyo}.
The OKA results will be
included in the averages for the form-factor parameters in a future update,
once the systematic uncertainties have been fully evaluated.

The dispersive parameterizations are affected by uncertainty in the
behavior of the $K\pi$ phase shifts in the high-energy region, which
in turn gives rise to uncertainties on constants in the expressions used
to fit the experimental data. The contributions of these uncertainties
to $\Lambda_+$ and $\ln C$ are evaluated by the experiments, and are
typically about $0.3\times10^{-3}$ for $\Lambda_+$ and $0.004$ for $\ln C$
(see, e.g., \cite{Bernard:2009zm,Abouzaid:2009ry}).
Since these contributions are common to all of the measurements, they are
removed from the input data before averaging and added back to the
uncertainties on the final results.
As shown in Fig.~\ref{fig:ffcomp},
until recently, the uncertainties from the parameterization were small on
the scale of the overall uncertainties for the individual measurements.
However, they are not small compared to the uncertainties for the average
values of $\Lambda_+$ and $\ln C$, and even the preliminary NA48/2
results are significantly affected. With correlations taken into account,
the uncertainties from the parameterization contribute about 0.05\% to the
uncertainties on the phase-space integrals~\cite{Abouzaid:2009ry}.
As measurements of the form-factor parameters become more precise,
it will become increasingly important to reduce the uncertainties arising
from the representation of the form factors. This underscores the
importance of preserving and presenting the experimental kinematic
distributions in a form that will allow refitting with improved
parameterizations when they become available.

Alternatively, the
$t$-dependence of the $K_{\ell3}$ form factors may someday soon be taken
from lattice QCD calculations. The ETM Collaboration has recently published
results from an $N_f=2+1+1$ calculation of the $K_{\ell3}$ form factors
in which synthetic data points representing the $t$-dependence of the
form factors were fit with the dispersive parameterization, giving
$\Lambda_+ = (24.22\pm1.16)\times10^{-3}$ and 
$\ln C = 0.1998(138)$ (with $\rho = +0.376$)~\cite{Carrasco:2016kpy}.
These values are in reasonable agreement with the experimental average; their
uncertainties are not too much larger than those for some of the
individual measurements contributing to the average. 

\section{Evaluation of $V_{us}$ and related tests}

\begin{table}
\begin{center}
\begin{tabular}{lcccccc}
\hline\hline
              &              &         & \multicolumn{4}{@{}c@{}}{\small Approx contrib to \% err} \\
Mode          & $V_{us}\,f_+(0)$ & \% err  & BR   & $\tau$ & $\Delta$ & $I$  \\ 
\hline
$K_{L\,e3}$    & 0.2163(6)  & 0.25    & 0.09 & 0.20   & 0.11     & 0.05 \\ 
$K_{L\,\mu3}$  & 0.2166(6)  & 0.28    & 0.15 & 0.18   & 0.11     & 0.06 \\
$K_{S\,e3}$    & 0.2155(13) & 0.61    & 0.60 & 0.02   & 0.11     & 0.05 \\
$K^\pm_{e3}$   & 0.2171(8)  & 0.36    & 0.27 & 0.06   & 0.22     & 0.05 \\
$K^\pm_{\mu3}$ & 0.2170(11) & 0.51    & 0.45 & 0.06   & 0.22     & 0.06 \\
\hline\hline
\end{tabular}
\end{center}
\caption{\label{tab:Vusf}
Values of $V_{us}\,f_+(0)$ from data for different decay modes, 
with breakdown of uncertainty from different sources: branching ratio 
measurements (BR), lifetime measurements ($\tau$), long-distance radiative
and isospin-breaking corrections ($\Delta$), and phase-space integrals
from form-factor parameters ($I$).}
\end{table}
The evaluations of $V_{us}\,f_+(0)$ for each of the five decay modes
($K_{L\,e3}$, $K_{L\,\mu3}$, $K_{S\,e3}$, $K^\pm_{e3}$, and $K^\pm_{\mu3}$)
are presented in Table~\ref{tab:Vusf}, with a breakdown of the 
uncertainties from different sources in each case. The most precise
values are from the $K_L$ decays, where the dominant uncertainty is from
$\tau_{K_L}$. For the other channels, the dominant uncertainties are
from the BR measurements. The uncertainties from the phase-space integrals
are insignificant, although uncertain knowledge of the form-factor
parameters may limit the precision of next-generation BR measurements
by entering into the acceptance calculations.
The five-channel average is $V_{us}\,f_+(0) = 0.21654(41)$ with  
$\chi^2/{\rm ndf} = 1.54/4$ $(P = 82\%)$. 

Since there are no new experimental inputs to the analysis since 2014,
the figures in Table~\ref{tab:Vusf} are unchanged, except for very small
changes to the values of $V_{us}\,f_+(0)$ for the $K^\pm$ decays.
These changes arise from 
the correction for strong isospin breaking used in this analysis,
$\Delta_{SU(2)} = (2.45\pm0.19)\%$, which was
updated using the 2016
$N_f = 2+1$ FLAG averages for the quark-mass ratios $Q = 22.5(6)(6)$
and $m_s/\hat{m} = 27.43(13)(27)$, with the isospin-limit meson masses
$M_K = 494.2(3)$ MeV and $M_\pi = 134.8(3)$ MeV~\cite{Aoki:2016frl}
(see~\cite{Cirigliano:2011ny} for discussion and notation).
Previous to the advent of precise lattice results for the light-quark
masses, the uncertainty on $\Delta_{SU(2)}$
was a leading contribution to the 
uncertainty on $V_{us}\,f_+(0)$ from the charged-kaon modes. As seen
from Table~\ref{tab:Vusf},
%it is still a significant source 
%of uncertainty for these modes,
it still is, and the value of $\Delta_{SU(2)}$ used here 
should be confirmed and improved upon.
There is continuing progress on the estimation of the light-quark masses.
For example, Colangelo {\it et al.}\ have performed a dispersion-relation
analysis
of the $\eta\to3\pi$ Dalitz plot making use of KLOE data for the charged mode,
obtaining $Q=22.0(7)$~\cite{Colangelo:2016jmc}, while the BMW Collaboration
has obtained the result $Q=23.4(6)$ from a lattice calculation with $N_f=2+1$
and partially quenched QED~\cite{Fodor:2016bgu}.
A systematic review of these and other results would be quite useful to
refine the value of $\Delta_{SU(2)}$.
Averaging the results for $V_{us}\,f_+(0)$ separately for neutral and
charged kaons gives
$V_{us}\,f_+(0) = 0.2163(5)$ for $K^0$ and $0.2224(7)$ for $K^\pm$,
with  
$\chi^2/{\rm ndf} = 0.75/3$ and a negligible correlation from
the use of the same form-factor parameters for the evaluation of the
phase-space integrals in either case.
Perfect equality of the uncorrected results for $V_{us}\,f_+(0)$ 
from charged and neutral modes would then
require $\Delta_{SU(2)} = 2.82(38)\%$.

A value for $f_+(0)$ is needed to obtain the value of $V_{us}$. 
The FLAG review provides separate recommended values for $N_f=2+1$ and
$N_f=2+1+1$. For 2016, the FLAG value for $N_f=2+1$,
$f_+(0)=0.9677(27)$, was updated with the addition of a new result from
RBC/UKQCD~\cite{Boyle:2015hfa}; the value for $N_f=2+1+1$,
$f_+(0)=0.9704(32)$, from \cite{Bazavov:2013maa}, is unchanged from 2014.
After the FLAG 2016 cutoff, the $N_f=2+1+1$ result
from \cite{Carrasco:2016kpy}, $f_+(0)=0.9709(46)$, was
published; as noted above, values for
$\Lambda_+$ and $\ln C$ were also obtained in this analysis.
In addition, two ongoing studies were
reported at Lattice 2016; one of these, the FNAL/MILC update of the
$N_f=2+1+1$ result quoted by FLAG, expects to obtain an overall precision
of $\sim$0.2\%~\cite{Gamiz:2016bpm}, such that in the near future, the
precision on $f_+(0)$ from the lattice will become competitive with the
precision of $V_{us}\,f_+(0)$ from experiment.

The test of CKM unitarity requires a value for $V_{ud}$.
A preliminary update of the survey of experimental data on
$0^+\to0^+$ $\beta$ decays from Hardy and Towner~\cite{Hardy:2014qxa}
was presented at this conference, giving $V_{ud} = 0.97420(21)$
\cite{Hardy}.
The update includes a few new
measurements, including BR and $Q_{\rm EC}$ measurements for $^{14}$O.
The world data set on $0^+\to0^+$ $\beta$ decays is very robust at this
point and additional measurements have small effects on $V_{ud}$, the
dominant uncertainty on which is from the calculation of the
short-distance radiative correction, $\Delta_R$.

\begin{table}
\begin{center}
\begin{tabular}{ccccc}
\hline\hline
\multicolumn{2}{@{}c@{}}{Choice of $f_+(0)$} & $V_{us}$ &
\multicolumn{2}{@{}c@{}}{$\Delta_{\rm CKM}$} \\
\hline
$N_f = 2+1$   & 0.9677(27) & 0.2238(8) & $-0.0009(5)$ & $-1.6\sigma$ \\
$N_f = 2+1+1$ & 0.9704(32) & 0.2231(9) & $-0.0011(6)$ & $-2.0\sigma$ \\
\hline\hline
\end{tabular}
\end{center}
\caption{\label{tab:Vus}
Results for $V_{us}$ and first-row unitarity test from $K_{\ell3}$ decays.}
\end{table}
The results for $V_{us}$ from $K_{\ell3}$ decays with $N_f=2+1$ and 
$N_f=2+1+1$ lattice values for $f_+(0)$ and the latest value of 
$V_{ud}$ are listed in Table~\ref{tab:Vus}. Because the $N_f=2+1$ and
$N_f=2+1+1$ lattice values for $f_+(0)$ obtained since 2012 are larger
than the pre-2012 values, which are mainly from $N_f=2$ simulations,
the agreement with the expectation from first-row unitarity at the 0.1\%
level obtained in~\cite{Antonelli:2010yf} is no longer observed: 
$\Delta_{\rm CKM}$ is different
from zero by $-1.6\sigma$ and $-2.0\sigma$ when the $N_f=2+1$ and  
$N_f=2+1+1$ results for $f_+(0)$ are used, respectively.

\sloppypar
Up to kinematic factors and long-distance electromagnetic corrections,
$\delta_{\rm EM}$, 
the ratio of the inner-bremsstrahlung-inclusive rates for $K^\pm_{\mu2}$ 
and $\pi^\pm_{\mu2}$ decays provides access to the quantity 
$V_{us}/V_{ud}\times f_{K^\pm}/f_{\pi^\pm}$.
The addition to the $K^\pm$ fit of the
${\rm BR}(K^\pm\to\pi^\pm\pi^+\pi^-)$ measurement from KLOE-2 in 2014 
slightly increased ${\rm BR}(K^\pm_{\mu2})$ and reduced its uncertainty
from 0.3\% to 0.2\%, leading to the result 
$V_{us}/V_{ud}\times f_{K^\pm}/f_{\pi^\pm} = 0.27599(37)$.
This result is obtained using $\delta_{\rm EM} = -0.0069(17)$, from the
chiral-perturbation theory analysis of~\cite{Cirigliano:2011tm}.
Note that the quantity $f_{K^\pm}/f_{\pi^\pm}$ includes the effects 
of strong isospin breaking:
$f_{K^\pm}/f_{\pi^\pm} \equiv f_{K}/f_{\pi} \times [1 + \delta_{SU(2)}]^{1/2}$.
The ETM Collaboration has recently obtained the value
$\delta_{\rm EM} + \delta_{SU(2)} = -0.0137(13)$
from an $N_f=2+1+1$ lattice simulation, to be compared with the
value from \cite{Cirigliano:2011tm},
$\delta_{\rm EM} + \delta_{SU(2)} = -0.0112(21)$.
The ETM result is preliminary and the uncertainty from the use of
the quenched QED approximation is not included, but it highlights
the potential role of the lattice for checking and improving on
corrections for electromagnetic effects.
The 2016 FLAG average of four complete and published $N_f = 2+1$ 
determinations of $f_{K^\pm}/f_{\pi^\pm}$ is 1.192(5), unchanged from
2014.
The $N_f = 2+1+1$ average, $f_{K^\pm}/f_{\pi^\pm} = 1.1933(29)$, is dominated by 
results from HPQCD~\cite{Dowdall:2013rya} and
Fermilab/MILC~\cite{Bazavov:2014wgs}
which were obtained using in part the same staggered-quark ensembles
generated by MILC.

\begin{table}
\begin{center}
\begin{tabular}{ccccc}
\hline\hline
\multicolumn{2}{@{}c@{}}{Choice of $f_{K^\pm}/f_{\pi^\pm}$} & $V_{us}/V_{ud}$ &
\multicolumn{2}{@{}c@{}}{$\Delta_{\rm CKM}$} \\
\hline
$N_f = 2+1$   & 1.192(5)   & 0.2315(10) & $-0.00004(59)$ & $-0.06\sigma$ \\
$N_f = 2+1+1$ & 1.1933(29) & 0.2313(6) & $-0.00015(48)$ & $-0.3\sigma$ \\
\hline\hline
\end{tabular}
\end{center}
\caption{\label{tab:Vusd}
Results for $V_{us}/V_{ud}$ and first-row unitarity test from 
$K^\pm_{\mu2}$ decays.}
\end{table}
The results for $V_{us}/V_{ud}$ and the first-row unitarity test 
from $K^\pm_{\mu2}$ decays obtained with the $N_f = 2+1$ and $N_f = 2+1+1$
lattice values for $f_{K^\pm}/f_{\pi^\pm}$ and $V_{ud}$ from
\cite{Hardy} are presented in Table~\ref{tab:Vusd}.
The agreement with first-row unitarity is better for the case of
$K^\pm_{\mu2}$ decays than for $K_{\ell3}$ decays.

\begin{figure}
\begin{center}
\includegraphics[width=0.3\textwidth]{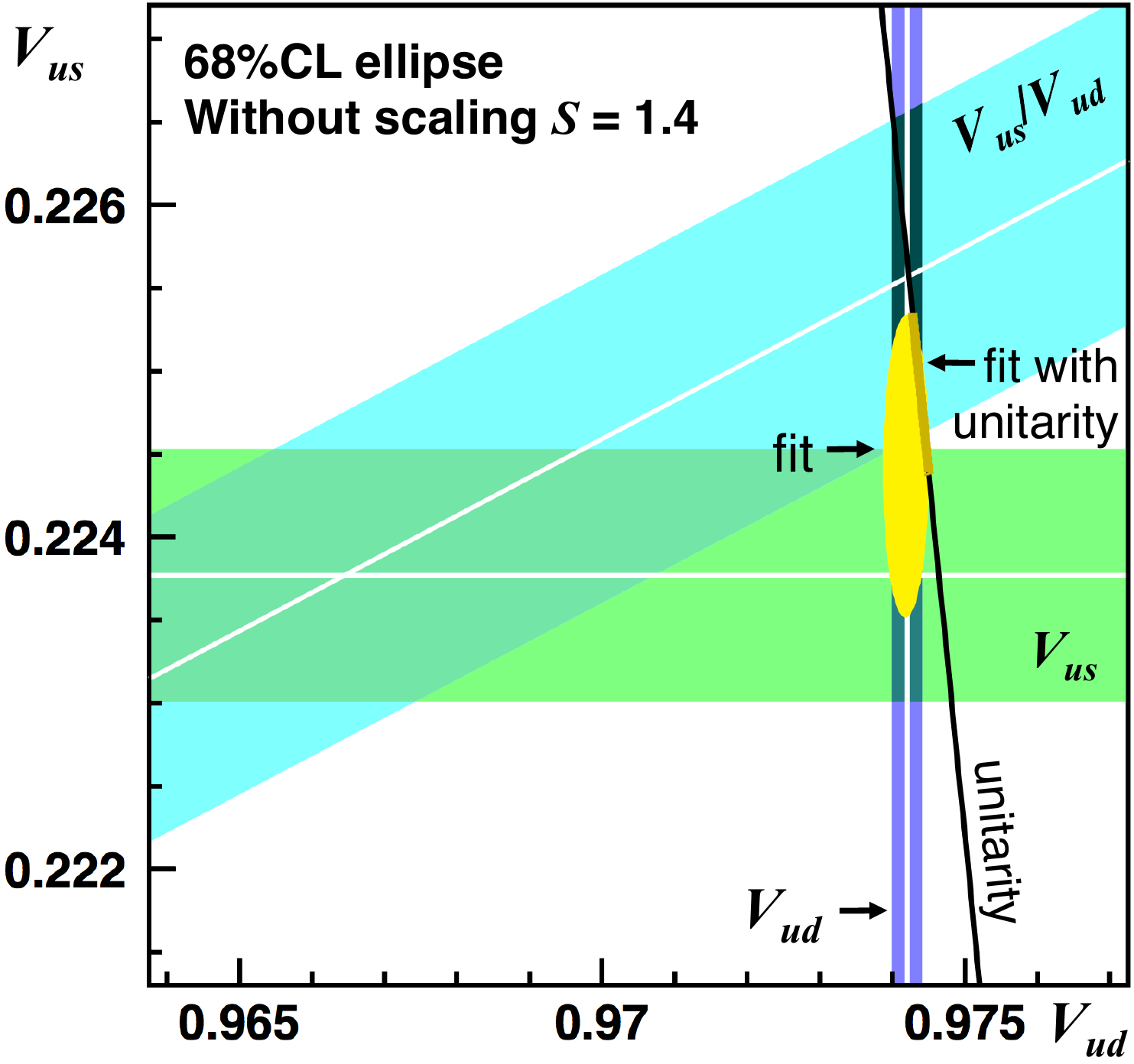}
\hspace{5mm}
\includegraphics[width=0.3\textwidth]{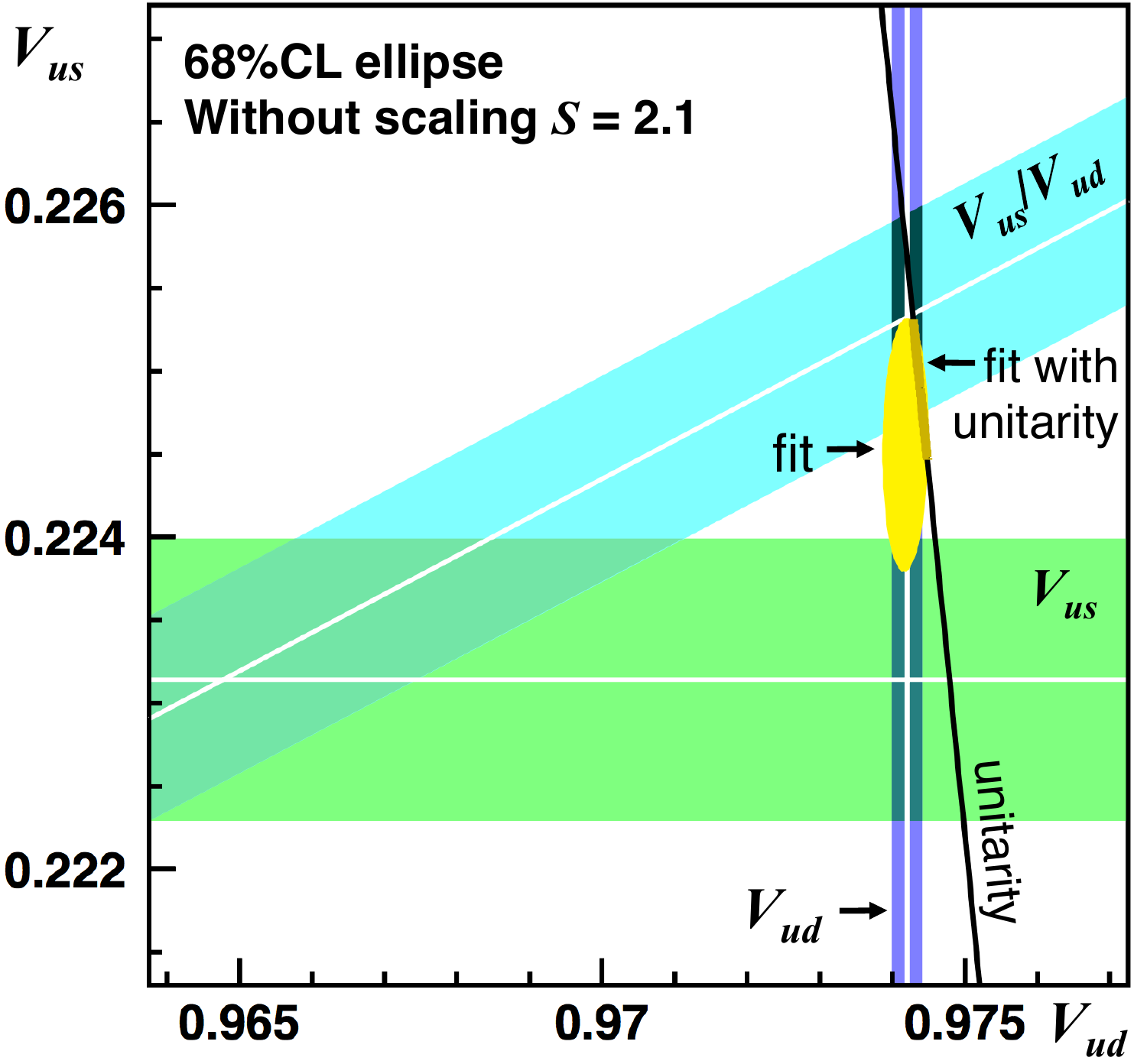}
\caption{Fits to $V_{ud}$ from $0^+\to0^+$ $\beta$ decays, $V_{us}$ from
$K_{\ell3}$ decays, and $V_{us}/V_{ud}$ from $K^\pm_{\mu2}$ decays,
with lattice values of $f_+(0)$ and $f_{K^\pm}/f_{\pi^\pm}$ from 
$N_f=2+1$ (left) and $N_f=2+1+1$ (right). The yellow 
ellipses indicate the 68\% confidence intervals in the plane of
($V_{ud}$, $V_{us}$) for the fits with with no constraints. The yellow 
line segments indicate the results obtained with the constraint 
$\Delta_{\rm CKM} = 0$.}
\label{fig:univers}
\end{center}
\end{figure} 
The values of $V_{ud}$ from $0^+\to0^+$ $\beta$ decays, $V_{us}$ from 
$K_{\ell3}$ decays, and $V_{us}/V_{ud}$ from $K^\pm_{\mu2}$ decays can be 
combined in a single fit, as illustrated in Fig.~\ref{fig:univers}.
The fit can be performed with or without
the unitarity constraint, $\Delta_{\rm CKM} = 0$. 
The unconstrained fits give $V_{us}=0.2244(6)$ with
$\Delta_{\rm CKM} = -0.0006(5)$
$(-1.2\sigma)$ for the analysis using $N_f = 2+1$ lattice results,
and $V_{us}=0.2246(5)$ with $\Delta_{\rm CKM} = -0.0005(5)$
$(-1.1\sigma)$ for the analysis using $N_f = 2+1+1$ results,
However, because of the change in the value of $V_{us}$ from $K_{\ell3}$
following from the increased value of $f_+(0)$ from the recent
generation of lattice results, the $K_{\ell3}$ and $K^\pm_{\mu2}$ results
are not as consistent as previously observed, and the uncertainties quoted
for $V_{us}$ do not include scale factors of 1.4 and 2.1
from the $\chi^2$ values of the respective fits.

\section{Outlook}

All of the post-2010 experimental results combined have a 
marginal effect on the results of the unitarity test, and good agreement
with unitarity is still observed for $K^\pm_{\mu2}$ decays. The question arises
as to whether hidden systematics in the $K_{\ell3}$ data and/or lattice
calculations are becoming important as the stated uncertainties shrink.
On the experimental side, the consistency of the fits to $K_{\ell3}$ 
rate data is creaky. Yet, the errors on the BRs from these fits are scaled
to reflect internal inconsistencies, and after this procedure, the
values of $V_{us}\,f_+(0)$ from $K_L$, $K_S$, and $K^\pm$ modes show
good agreement. There is
also a fair amount of redundancy in the $K_{\ell3}$ data set, so adding or
eliminating single measurements doesn't change the results for 
$V_{us}\,f_+(0)$ by much.

Fortunately, a new generation of experiments holds forth the promise
of new results that may help to clarify this situation.
Since the dominant errors are systematic for most measurements,
new, high-statistics data would be principally helpful to the extent
that it provides samples for detailed systematic studies.
Any of the following
experiments could potentially contribute:
\begin{itemize}

\item {\bf NA62} The successor experiment to NA48/2, NA62~\cite{NA62:2017rwk}
  aims to measure ${\rm BR}(K^+\to\pi^+\nu\bar{\nu})$ to $\sim$10\%.
  When running at low intensity, NA62 collects on the order of 1M
  $K_{e3}$ decays per week.
  Relative to NA48/2, NA62 has better $\pi/\mu$ separation
  and full beam tracking to help with the reconstruction of $t$ for
  form-factor measurements, though the presence of additional material
  upstream of the calorimeter may partially offset these advantages.
  NA48/2 itself has yet to finalize its preliminary measurement of
  the $K_{\ell3}$ form-factor parameters,
  and data acquired by NA62 2007 with the NA48/2 setup could also be
  analyzed. 
\item {\bf OKA} A fixed-target experiment at the U-70 synchrotron in
  Protvino, OKA is a successor to ISTRA+, installed in a new beamline
  with an RF-separated $K^+$ beam. OKA can measure $K^+$ BRs, and as noted
  above, has an analysis of the $K_{e3}$ form factor in progress. In runs
  from 2010 to 2013, OKA collected $\sim$17M $K_{e3}$ events; the experiment
  took data again in late 2016~\cite{Obraztsov:2017oyo}.
\item {\bf KLOE-2} Like its predecessor, KLOE-2 can measure the full suite
  of observables for $V_{us}$, including BRs, form-factor parameters, and,
  importantly, lifetimes for the $K^\pm$, $K_L$, and $K_s$.
  KLOE-2 started taking data at the end of 2014 and is expected to
  collect 5~fb$^{-1}$ total by the end of 2017~\cite{Passeri:2017ukw}.
  In addition,
  2~fb$^{-1}$ of KLOE data have yet to be fully exploited. The original
  KLOE $V_{us}$ analyses were based on about $0.4~$fb$^{-1}$ of data.
  KLOE measurements related to $V_{us}$ are limited by systematics, but
  the high statistics KLOE-2 data should allow a precise measurement
  of the BRs for $K_{S\,\ell3}$, particularly $K_{S\,e3}$.
  Because $\tau_{K_S}$ is known to 0.04\%, $K_{S\,e3}$
  could be the best channel for the determination of $V_{us}$.
  The new data might also allow KLOE-2 to improve on KLOE results for the
  form-factor parameters (including those for $K^\pm$) and on the
  $K_L$ lifetime.
\item {\bf LHCb} $10^{13}$ $K_S$/fb$^{-1}$ are produced inside the LHCb
  acceptance. The recent limit on ${\rm BR}(K_S\to\mu\mu)$~\cite{Aaij:2012rt}
  (and preliminary update~\cite{RamosPernas:2017xdi}) demonstrates LHCb's
  capability to measure $K_S$ decays to muons.
  LHCb might be able to measure BR($K_{S\,\mu3}$), although relative to
  $K_S\to\mu\mu$ there are additional difficulties because of the
  incomplete reconstruction of the final state, and the presence of
  $K_{L\,\mu3}$ would make lifetime analysis necessary. There is a possible
  limitation from the trigger, but with a dedicated high-level trigger line,
  this might be overcome.
  Like $K_{S\,e3}$, $K_{S\,\mu3}$ offers high sensitivity because 
  $\tau_{K_S}$ is precisely known, though at LHCb this fact may have to be
  exploited for $K_S/K_L$ separation. On the other hand, BR($K_{S\,\mu3}$)
  has never been measured and would provide a new channel for the
  measurement of $V_{us}$.
\item {\bf TREK} TREK is designed to measure the $T$-violating transverse
  muon polarization in $K_{\mu3}$ decay. A first, low-intensity phase of
  the experiment took data in 2015 with the goal of measuring
  ${\rm BR}(K_{e2})/{\rm BR}(K_{\mu2})$ to within 0.25\%, among
  others~\cite{Bianchin:2016vds}.
  The experiment makes use of an upgraded KEK-246 setup, moved to J-PARC:
  $K^+$s are stopped in an active target surrounded by a toroidal
  spectrometer, with EM calorimetry and redundant $e/\mu$ identification.
  Since KEK-246 measured
  ${\rm BR}(K_{\mu3})/{\rm BR}(K_{e3})$, %\cite{Horie:2001th},
  TREK should also be able to make BR measurements of interest for $V_{us}$.
\end{itemize}
In conclusion, there are good prospects for a new round of experimental
results to reduce the uncertainty on $V_{us}\,f_+(0)$ from 0.18\% at
present to $\sim$0.12\% within 5 years. Perhaps more important than
reducing the uncertainty on $V_{us}$ per se, this should allow a critical
test of the consistency of the results for $K_{\ell3}$ and better
comparison with the result from $K^\pm_{\mu2}$.

\acknowledgments

I would like to warmly thank F.~Dettori, V.~Duk, E.~G\'amiz,
V.~Obraztsov, M.~Rotondo, S.~Simula, R.~Vazquez Gomez, and O.~Yushchenko
for useful discussions. I am especially grateful to E.~Passemar for
detailed discussions regarding isospin breaking in $K_{\ell3}$ decays,
for checking the calculation of $\Delta_{SU(2)}$, and for kindly reading
the manuscript.

\end{document}